\documentclass[12pt]{article}
\usepackage{a4wide}
\usepackage{amssymb}
\begin{document}
{\renewcommand{\thefootnote}{\fnsymbol{footnote}}
%\hfill  IGC--yy/m--n\\
\medskip
\begin{center}
{\LARGE  Deformed General Relativity }\\
\vspace{1.5em}
Martin Bojowald\footnote{e-mail address: {\tt bojowald@gravity.psu.edu}}
and George M.~Paily\footnote{e-mail address: {\tt gmpaily@phys.psu.edu}}
\\
\vspace{0.5em}
Institute for Gravitation and the Cosmos,\\
The Pennsylvania State
University,\\
104 Davey Lab, University Park, PA 16802, USA\\
\vspace{1.5em}
\end{center}
}

\setcounter{footnote}{0}

\newcommand*{\R}{{\mathbb R}}
\newcommand*{\N}{{\mathbb N}}
\newcommand*{\Z}{{\mathbb Z}}
\newcommand*{\Q}{{\mathbb Q}}
\newcommand*{\C}{{\mathbb C}}

\begin{abstract}
  Deformed special relativity is embedded in deformed general relativity using
  the methods of canonical relativity and loop quantum gravity. Phase-space
  dependent deformations of symmetry algebras then appear, which in some
  regimes can be rewritten as non-linear Poincar\'e algebras with
  momentum-dependent deformations of commutators between boosts and time
  translations. In contrast to deformed special relativity, the deformations
  are derived for generators with an unambiguous physical role, following from
  the relationship between canonical constraints of gravity with stress-energy
  components. The original deformation does not appear in momentum space and
  does not give rise to non-locality issues or problems with macroscopic
  objects. Contact with deformed special relativity may help to test loop
  quantum gravity or restrict its quantization ambiguities.
\end{abstract}

\section{Introduction}

A quantum theory of gravity combines the fundamental constants of nature $G$
and $\hbar$, characteristic of the ingredients of general relativity and
quantum mechanics. It should therefore assign a specific role to the Planck
length $\ell_{\rm P}=\sqrt{G\hbar}$, the Planck mass $m_{\rm
  P}=\hbar/\ell_{\rm P}$ or the Planck density $\rho_{\rm P}=m_{\rm
  P}/\ell_{\rm P}^3$ beyond what one may expect on purely dimensional
grounds. One suggestion that is often made is the presence of an invariant
length, $\ell_{\rm P}$, on the same footing as the invariant speed of light
$c$ in special and general relativity. Even more specifically, $\ell_{\rm P}$
may pose a lower limit to distances, or $\rho_{\rm P}$ an upper limit to
densities, or $m_{\rm P}$ an upper limit to masses and energies. With these
assumptions, especially the last one, one can be led to different versions of
deformed special relativity \cite{DSR1,DSR2,DSR,DSRSoccer}, based on
deformations of the Poincar\'e algebra so that a second invariant constant is
introduced.

While an invariant shortest distance or largest density may sound natural in
quantum gravity, it is by no means implied just by the fact that $G$ and
$\hbar$ both appear in the theory. Setting aside the question of bounds, it is
not even clear whether there should be an invariant distance or mass. While
the ingredients $G$ and $\hbar$ of $\ell_{\rm P}$ and $m_{\rm P}$, and so the
Planck quantities themselves, must be
invariant under transformations of reference frames, the question of invariant
distances or masses depends on what role $\ell_{\rm P}$ and $m_{\rm P}$ play
for physical observables. The question of observables or measurement
procedures is complicated in any combination of quantum physics with general
relativity, and therefore detailed knowledge (rather than just dimensional
expectations) is required before these questions can be answered.

What gives substance to the claims of deformed special relativity is the fact
that symmetries and their quantum realizations present some of the most
fundamental concepts in physics. The mathematical rigidity of possible
deformations of the Poincar\'e algebra by quantum corrections or other effects
allows interesting tests of the current understanding of quantum gravity in
general terms, or, if such effects are derived from one of the candidate
theories, means to compare the different, usually disparate approaches. In
this article, we take this viewpoint and have a general look at canonical
quantum gravity.\footnote{Our discussions and results are different from
  earlier attempts to derive deformed special relativity or Lorentz violations
  from quantum gravity. In \cite{DeformedQG}, $2+1$-dimensional models have
  given rise to deformed Poincar\'e algebras, but the key property
  (non-vanishing Poisson brackets of gravitational connection components) is
  not necessarily realized in $3+1$ dimensions. Another suggestion made in
  \cite{DeformedQG} for $3+1$ dimensions makes use of Chern--Simons-like
  boundary terms of Plebanski actions, whose algebra turns out to be deformed
  \cite{TQFTQG}. We will make use of boundary observables as well, but already
  the bulk terms of canonical gauge generators will obey a deformed algebra.
  More recently, in \cite{LorentzBianchi} methods related to quantum field
  theory on a modified space-time background have been used, but no clear
  deformation or violation effects have been found. Moreover, the latter
  analysis ignored a consistent treatment of quantum space-time structures,
  the key ingredient used here.}

\section{From Poincar\'e transformations to hypersurface deformations}

When combined with gravity, space-time described by special relativity is too
limited. One should rather use general relativity and its richer structure of
arbitrary coordinate transformations. Deformed special relativity, in which
general covariance is not realized but a non-zero gravitational constant (and
in some arguments gravitational phenomena such as black holes) is
assumed can be considered only as a limit. But it is not clear whether there
is a consistent relativistic procedure that does away with general covariance
but still keeps the gravitational constant as a fundamental parameter. For
this reason, we propose to ask the question of possible deformations for the
symmetries underlying general relativity. In algebraic form, we go from the
well-known Poincar\'e relations\footnote{We will always use commutators of
  classical type, computed as Poisson brackets $\{\cdot,\cdot\}$ of
  phase-space functions. A quantum analog of operators would then read
  $[\cdot,\cdot]=i\hbar\{\cdot,\cdot\}$. Although we will introduce quantum
  effects as crucial ingredients for deformations, they will be computed from
  effective equations. We therefore write $\{\cdot,\cdot\}$ instead of
  $[\cdot,\cdot]$ to avoid the impression that we are requantizing terms of
  effective equations.} 
\begin{eqnarray}
  \{P_{\mu},P_{\nu}\} &=& 0\, \label{Poincare1}\\
  {} \{M_{\mu\nu},P_{\rho}\} &=& \eta_{\mu\rho}P_{\nu}-
  \eta_{\nu\rho}P_{\mu} \label{Poincare2}\\ 
  {} \{M_{\mu\nu},M_{\rho\sigma}\} &=& \eta_{\mu\rho}M_{\nu\sigma}-
  \eta_{\mu\sigma}M_{\nu\rho}- \eta_{\nu\rho}M_{\mu\sigma}+
  \eta_{\nu\sigma}M_{\mu\rho}  \label{Poincare3}
\end{eqnarray}
to the much more unwieldy algebra of hypersurface deformations in space-time.

\subsection{Hypersurface-deformation algebra}

According to Dirac \cite{DiracHamGR}, a canonical field theory on space-time
foliated by equal-time slices is generally covariant if it is invariant under
the hypersurface-deformation algebra
\begin{eqnarray}
 \{D[M^a],D[N^a]\}&=& D[{\cal L}_{N^b}M^a] \label{DD} \\
 \{H[M],D[N^a]\}&=& H[{\cal L}_{N^b}M] \label{HD}\\
 \{H[M],H[N]\}&=& D[h^{ab}(M\nabla_bN-N\nabla_bM)] \label{HH}
\end{eqnarray}
whose generators $D[N^a]$ and $H[N]$ depend on shift vector fields $N^a$ and
lapse functions $N$ on the spatial slices. (For an introduction to methods and
properties of canonical gravity used in this paper we refer to \cite{CUP}.)
Also the metric $h_{ab}$ on spatial slices, or its inverse $h^{ab}$, appears
in the structure functions. If $D[N^a]$ and $H[N]$ are realized as gauge
generators, an infinitesimal space-time diffeomorphism along a vector field
$\xi^{\mu}$ is represented by the gauge transformation
$\delta_{\epsilon^{\mu}}f= \{f,H[\epsilon]+D[\epsilon^a]\}$ on phase-space
functions, with $\epsilon=N\xi^0$ and $\epsilon^a=\xi^a+N^a\xi^0$
\cite{LapseGauge}. The hypersurface-deformation algebra therefore describes
general covariance, just as the Poincar\'e algebra describes the symmetries of
special relativity.

The generators $D[N^a]$ and $H[N]$ consist of bulk terms which vanish as the
canonical constraints, and spatial boundary terms if they are computed for
finite regions or in space-times with specific asymptotic fall-off
conditions. Boundary terms are not required to vanish and provide energy and
(angular) momentum observables, for finite regions of Brown--York type
\cite{BrownYork} and for asymptotic regions of ADM type
\cite{ReggeTeitelboim}. The generators are therefore physically related to
those of the Poincar\'e algebra, just as the transformations are geometrically
related. In both views, however, the freedom in the algebra is much larger for
hypersurface deformations, which are not required to be linear, and for their
generators, whose physical expressions as energy and momentum refer to a large
set of observers in different states of motion depending on which bounded or
asymptotic region is chosen.

The hypersurface-deformation algebra differs from the Poincar\'e algebra in
several important respects, not only in the fact that it is much larger and in
fact infinite-dimensional. While both algebras depend on the metric, these
coefficients in the case of the Poincar\'e algebra (\ref{Poincare2}) and
(\ref{Poincare3}) are constants because they just refer to Minkowski
space-time. The spatial metric in (\ref{HH}), on the other hand, in general
depends on the position and is a spatial tensor. The Minkowski metric in the
Poincar\'e algebra determines structure constants; the spatial metric on a
slice in a curved space-time used in the hypersurface-deformation algebra
determines structure functions. The hypersurface-deformation algebra is not a
Lie algebra, but a Lie algebroid \cite{ConsAlgebroid}. Its deformations, in
constrast to the Poincar\'e algebra, have not been studied systematically, and
therefore it presents an interesting, more-general object in the context of
deformed relativity. By its relation to general covariance, it automatically
incorporates gravity.

If there are reasons to believe that the Poincar\'e algebra is deformed, there
should be a corresponding deformed version of the hypersurface-deformation
algebra, to make sure that the gravitational force can be described
consistently under the deformation. Vice versa, if there is a deformation of
the hypersurface-deformation algebra,\footnote{In phrases like this, we use
  ``deformation'' in two different meanings. The context makes it clear which
  is implied.} it entails a deformation of the Poincar\'e algebra. While it is
difficult to embed deformed Poincar\'e algebras in a hypersurface-deformation
algebra, deriving a deformed Poincar\'e algebra from a deformed
hypersurface-deformation algebra can be accomplished by restricting the
algebra to linear functions $N^a$ and $N$ in a given set of coordinates,
together with Euclidean spatial slices such that $h_{ab}=\delta_{ab}$.
Choosing
\[
 N(x)=\Delta t+v_a x^a\quad,\quad
N^a(x)= \Delta x^a+R^a_b x^b
\]
relates hypersurface deformations to Poincar\'e transformations with time
translation $\Delta t$, spatial translations $\Delta x^a$, boosts by $v_a$
and spatial rotations with matrices $R^a_b$. We call the resulting
Poincar\'e-type algebra the {\em linear limit} of the (deformed)
hypersurface-deformation algebra we start with.

In the presence of deformed algebras, the corresponding space-time structure
differs from the classical one: gauge transformations do not agree with Lie
derivatives, and any dynamics consistent with a deformed algebra differs from
general relativity \cite{Regained,LagrangianRegained,Action}.\footnote{There
  is no ``effective line element'' in such a situation: Gauge transformations
  of $h_{ab}$ do not match with coordinate transformations of ${\rm d}x^a$ to
  give an invariant ${\rm d}s^2$. Nevertheless, all observables of interest
  can be computed by canonical methods. Comparisons with deformed Poincar\'e
  algebras in the linear limit may suggest corresponding quantum space-time
  models, such as $\kappa$-Minkowski. We will come back to this question at
  the end of this article.}  The identification of a deformed Poincar\'e
algebra as a restriction of the hypersurface-deformation algebra with linear
$N^a$ and $N$ may therefore seem ambiguous. However, even without reference to
classical space-times, the linear limit is distinguished. Linear $N$ and
$N^a$ in (\ref{DD})--(\ref{HHbeta}) lead to the only closed subalgebra if
$h^{ab}$ is constant. Therefore, if there is a deformed Poincar\'e algebra, it
can only be the linear limit of the hypersurface-deformation
algebra. 

For the linear limit to be meaningful, we assume, as always in
special-relativistic situations, that all energies and momenta involved are
sufficiently small and that their back-reaction on space-time can be
ignored. Using a constant $h_{ab}=\delta_{ab}$ is then justified. If
back-reaction cannot be ignored, there is simply no special-relativity limit
of the theory.

\subsection{Deformations}

Several examples of deformed hypersurface-deformation algebras have been found
in loop quantum gravity, using effective methods and operator
calculations. All these deformations leave the $D$-relations (\ref{DD}) and
(\ref{HD}) unchanged, while (\ref{HH}) is modified to 
\begin{equation}
  \{H[M],H[N]\}= D[\beta h^{ab}(M\nabla_bN-N\nabla_bM)] \label{HHbeta}
\end{equation}
with a phase-function $\beta\not=1$ that may depend on the spatial metric or
extrinsic curvature. Loop quantum gravity therefore confirms the expectation
that quantum geometry should lead to deformations of symmetry algebras of
space-time. In fact, no undeformed consistent version of symmetries at this
quantum level has been found. 

There is a broad consensus in loop quantum gravity that {\em off-shell}
constrained algebras must be deformed if quantum-geometry effects of the
theory are included. (See \cite{Action} for a detailed list of models.) The
first such deformations have been found by effective methods in models of
perturbative inhomogeneity \cite{ConstraintAlgebra} and in spherical symmetry
\cite{LTBII}, in both cases using inverse-triad corrections
\cite{QSDV,InvScale}. A second type of corrections, holonomy corrections, has
been implemented consistently in the same type of models
\cite{JR,ScalarHol}. Analogous deformations appear in operator calculations of
the constraint algebra for $2+1$-dimensional models, with holonomy corrections
\cite{ThreeDeform} and inverse-triad corrections
\cite{TwoPlusOneDef,TwoPlusOneDef2,AnoFreeWeak}. (An especially striking
feature of holonomy corrections is that they trigger signature change at high
density \cite{Action,SigChange}.)

In all cases, the algebra is deformed in the same way, with characteristic
functions $\beta$ depending on extrinsic curvature for holonomy corrections
and on the spatial metric for inverse-triad corrections. Unmodified space-time
structures appear only in cases in which the classical structure is
presupposed by fixing the gauge before quantization, a procedure which in
cosmology (and elsewhere) is known to lead to incorrect results. Deformed
hypersurface-deformation algebras are therefore an unavoidable consequence of
the quantization steps undertaken in loop quantum gravity, in particular the
use of holonomies as basic operators \cite{Rov,ThomasRev}. Loop quantum
gravity leads to deformed space-time structures and to deformed general
relativity in a semiclassical limit.

All quantum corrections are state-dependent and must therefore be
parameterized suitably, given that knowledge of quantum-gravity states is
limited.  Inverse-triad corrections in loop quantum gravity lead to a
deformation function $\beta$ depending on the size of discrete plaquettes
relative to the Planck area, and therefore on the spatial metric. Holonomy
corrections depend on the momentum of the spatial metric, related to extrinsic
curvature or the time derivative of the spatial metric.\footnote{Holonomy
  corrections are often claimed to be uniquely determined by classical
  parameters rather than states, especially in cosmology where they are
  supposed to depend only on the classical density divided by the Planck
  density. However, the Planck density in this case is chosen ad-hoc, and in
  general must be replaced by the density of discrete patches, a parameter of
  the quantum-gravity state. Also holonomy corrections therefore depend on the
  quantum-gravity state and must be parameterized. Regarding the number of
  parameters, there is no difference between holonomy corrections and
  inverse-triad corrections.} Without using detailed expressions which can be
derived in loop quantum gravity, one can easily expect corrections of these
two types. Inverse-triad corrections incorporate implications of discrete
space, while holonomy corrections implement additional curvature required to
embed discrete space in quantum space-time. In addition, there are corrections
from standard quantum fluctuations of the metric, which are more difficult to
compute in loop quantum gravity and have not yet been formulated in a
consistent form of hypersurface deformations. (Their main effect is to
introduce higher time derivatives \cite{EffAc,HigherTime}. These corrections
are therefore close relatives of higher-curvature terms, which do not modify
the hypersurface-deformation algebra \cite{HigherCurvHam}.)

When $\beta$ depends on the metric or extrinsic curvature $K_{ab}$, our
previous arguments about the Poincar\'e limit as the linear restriction of the
hypersurface-deformation algebra still apply. Linear $N^a$ and $N$ lead to a
unique subalgebra if $h^{ab}$ and $K_{ab}$ (and therefore $\beta$) are
spatially constant. In strong quantum regimes, curvature is large and the
fields could not be assumed constant. Under these conditions one does not
expect a Poincar\'e algebra to capture space-time properties. A Poincar\'e
description should be valid when quantum effects are not strong and the energy
observables are sufficiently small so that back-reaction on space-time can be
ignored. Under these conditions, it is safe to assume that the gravitational
fields are constant in regions of interest. A distinguished (deformed)
Poincar\'e algebra then follows from the hypersurface-deformation algebra.

However, the deformation is not of the form of non-linear Lie brackets (as
part of Hopf algebras) because the modified structure function depends on the
phase-space variables in a modified way but does not introduce non-linearities
in the generators $D$ and $H$. Going from Poincar\'e's Lie algebra to Dirac's
Lie algebroid with structure functions has led to a new option for
deformations, one not considered before. The expectations of deformed special
relativity are therefore not realized, at least not in general. Nevertheless,
there are deformations of underlying symmetries which one can try to test as
proposed and extensively analyzed in the context of deformed special
relativity. Moreover, as we show in what follows, there are regimes in which
one can relate phase-space dependent deformations to non-linear algebraic
structures.

\subsection{Holonomy corrections and energy-dependent deformations}

For holonomy corrections, one can, in certain regimes, relate
background-dependent deformations in (\ref{HHbeta}) to non-linear algebraic
relations. In this case, $\beta$ depends on extrinsic-curvature components. It
has not been possible yet to formulate full non-local holonomies consistently,
integrating the connection along curves in arbitrary directions. However, in
spherically symmetric models one can implement holonomies along curves on
spherical orbits, which then depend on an extrinsic-curvature component
$K_{\varphi}$ in an angular direction. This component is related to the orbit
area $A(x)$ at radial coordinate $x$ by $K_{\varphi}=-N^{-1}{\rm
  d}\sqrt{A}/{\rm d}t$. The spatial derivative of $\sqrt{A}$, $k={\rm
  d}\sqrt{A}/{\rm d}x$, is proportional to the trace of the extrinsic
curvature of the orbit 2-sphere in 3-dimensional space.

Before we relate these quantities to observables, we recall features of
spherically symmetric models in connection variables
\cite{SphSymm,SphSymmHam}. With components of a densitized triad
\begin{equation} \label{E} 
E^a_i\tau^i\frac{\partial}{\partial
    x^a}=E^x(x)\tau_3\sin\vartheta\frac{\partial}{\partial x}+
  E^{\varphi}(x)\tau_1\sin\vartheta\frac{\partial}{\partial\vartheta}+
  E^{\varphi}(x)\tau_2\frac{\partial}{\partial\varphi}\,,
\end{equation}
whose internal space is written as the Lie algebra su(2) with generators
$\tau_j=-\frac{1}{2}i\sigma_j$ in terms of Pauli matrices, the spatial metric
is
\begin{equation}
{\rm d}s^2=\frac{(E^{\varphi}(x))^2}{|E^x(x)|}{\rm d} x^{2}
+|E^{x}(x)|{\rm d}\Omega^{2}\,.
\end{equation}
(As a consequence, $A(x)=|E^x(x)|$.)  The components $E^x$ and $E^{\varphi}$
are canonically conjugate to extrinsic-curvature components $K_x$ and
$K_{\varphi}$ in
\begin{equation} \label{K}
 K_a^i\tau_i{\rm d}x^a=K_x(x)\tau_3{\rm d} x+K_{\varphi}(x)\tau_1{\rm d}\vartheta+
K_{\varphi}(x)\tau_2\sin\vartheta{\rm d}\varphi
\end{equation}
but not to connection components as in the full theory. (Note however, that
$K_x$ is simply a gauge-invariant version of the corresponding connection
component $A_x$ with respect to a remnant U(1)-gauge freedom of internal
rotations fixing $\tau_3$.) Using the general relations between extrinsic
curvature and time derivatives of the spatial metric, one computes
\begin{equation}
K_{\varphi}=-\frac{1}{2N\sqrt{|E^{x}|}} \frac{{\rm d}E^x}{{\rm d}t}
\quad\mbox{and}\quad
K_{x}=-\frac{1}{N\sqrt{|E^{x}|}}\left(\frac{{\rm d}{E}^{\varphi}}{{\rm d}t}-
\frac{E^{\varphi}}{2E^{x}} \frac{{\rm d}E^x}{{\rm d}t}\right)\,.
\end{equation}
These relations follow from equations of motion generated by the Hamiltonian
constraint and are modified if quantum-geometry corrections of loop quantum
gravity are included. The following considerations are independent of these
relations.

Holonomy corrections which replace $K_{\varphi}$ by $\sin(\delta
K_{\varphi})/\delta$ (or some other function with related properties) in the
Hamiltonian constraint, with a parameter $\delta$ that could depend on the
triad components, especially $E^x$, can be implemented consistently
\cite{JR}. They imply a deformed hypersurface-deformation algebra
(\ref{HHbeta}) with $\beta(K_{\phi})=\cos(2\delta K_{\varphi})$.\footnote{The
  phase-space function $K_{\varphi}(x)$ is evaluated at the fixed boundary
  used to define observables, so that no position dependence results.} More
generally, if $K_{\varphi}$ is replaced by a function $F(K_{\varphi})$,
$\beta(K_{\varphi})= \frac{1}{2} {\rm d}^2F^2/{\rm d}K_{\varphi}^2$
\cite{JR}. The deformation depends on phase-space variables rather than
algebra generators. However, for a specific choice of observers the
phase-space variables involved can be related to observables that play the
role of algebra generators in the linear limit.

Extrinsic-curvature components determine observables of Brown--York
\cite{BrownYork} or ADM \cite{ReggeTeitelboim} type: (angular) momentum
\begin{equation} \label{BYJ}
 P= 2\int_{\partial\Sigma}{\rm d}^2z v_b(r_ap^{ab}-
  \bar{r}_a\bar{p}^{ab})
\end{equation}
in direction $v^a$, measured by an observer who watches the spatial region
$\Sigma$. (A contribution from a reference metric with barred quantities is
subtracted to ensure that energy and momentum vanish in Minkowski space-time.)
The integrand depends on the co-normal $r_a$ of the boundary of $\Sigma$ and
the gravitational momentum
\begin{equation}
 p^{ab}(x) =
 \frac{\sqrt{\det h}}{16\pi G} (K^{ab}-K^c_ch^{ab})
\end{equation}
canonically conjugate to the spatial metric $h_{ab}$. In spherical symmetry,
we choose a spherical surface $\partial\Sigma$ of constant $x$, and compute
the gravitational momentum using the tensor
\begin{equation}
  K_{ab}{\rm d}x^a\otimes {\rm d}x^b= 
 K_x \frac{E^{\varphi}}{\sqrt{|E^x|}} {\rm d}x\otimes{\rm d}x+
  K_{\varphi} \sqrt{|E^x|} \left({\rm d}\vartheta\otimes{\rm d}\vartheta+
    \sin^2\vartheta {\rm d}\varphi\otimes{\rm d}\varphi\right)
\end{equation}
and its trace
\begin{equation}
 K^c_c= K_x \frac{\sqrt{|E^x|}}{E^{\varphi}}+
 2\frac{K_{\varphi}}{\sqrt{|E^x|}}
\end{equation}
(using $K_{ab}=K_a^ie_b^i$ with $E^a_i=e^a_i|\det e_b^j|$).
We find that the radial component of the gravitational momentum is
proportional to $K_{\varphi}$ and independent of $K_x$:
\begin{equation}
 p^{xx}= -\frac{1}{8\pi G} K_{\varphi} \frac{|E^x|}{E^{\varphi}}
 \sin\vartheta\,.
\end{equation}

The radial component $p^{xx}$ of the gravitational momentum appears in the
linear Brown--York momentum (\ref{BYJ}) with $v^a=(\partial/\partial x)^a$. We
compute
\[
 P_x= \frac{8\pi g_{xx}p^{xx}}{E^{\varphi}\sqrt{|E^x|}}= 
-\frac{1}{G} \frac{K_{\varphi}}{\sqrt{|E^x|}}\,.
\]
If $\delta\propto |E^x|^{-1/2}$ (corresponding, in the language of holonomy
corrections, to lattice refinement \cite{InhomLattice,CosConst} with sites of
constant size, or a site number per spherical orbit proportional to the orbit
area), the combination of variables is exactly what appears in the deformation
function $\beta=\cos(2\delta K_{\varphi})$. The linear limit of the
hypersurface-deformation algebra therefore suggests that the commutator
between a boost $B_x$ and a time translation $P_0$ is deformed by a function
depending on the spatial momentum $P_x$, of the form
\begin{equation}\label{BP}
 \{B_x,P_0\}= \cos(\lambda P_x) P_x
\end{equation}
with a constant $\lambda$.

The Brown-York energy is
\begin{equation}
 E=-\frac{1}{8\pi G}\int_{\partial\Sigma}{\rm d}^2z
N(\sqrt{\det\sigma}\:k-\sqrt{\det\bar{\sigma}}\:\bar{k})
\end{equation}
with the induced spatial metric $\sigma_{ab}$ on $\partial\Sigma$ and the
trace $k$ of extrinsic curvature of $\partial\Sigma$ in space, related to
$\partial E^x/\partial x$ for a spherical $\partial\Sigma$. This function does
not appear in holonomy corrections, and inverse-triad corrections depend on
$E^x$ integrated over elementary plaquettes of a discrete state (so-called
fluxes) rather than its spatial derivative. If a derivative expansion of the
non-local fluxes is used, derivatives of $E^x$ and therefore $k$ will appear,
relating the corresponding algebra deformation to the Brown--York
energy. However, the relation is not as direct as in the case of holonomy
corrections and the radial momentum.

\section{Comparison with deformed special relativity}

Different realizations of deformed Poincar\'e algebras are possible, depending
on the choice of generators. Moreover, one can remove deformations by
non-linear transformations of the generators, giving rise to the question of
which expression of the generators should be physically preferred. In fact, it
is the choice of generators that determines whether the algebra is
deformed. In our case, using the hypersurface-deformation algebra, the
generators have unambiguous physical meaning via the boundary terms used
above. Also the bulk terms, taken for matter contributions to the gauge
generators which obey the same algebra as the full generators in the absence
of derivative couplings, have a clear meaning as the matter energy density
\begin{equation} \label{rhoE}
 \rho_{\rm E}= \frac{H_{\rm matter}[N]}{N\sqrt{\det h}}
\end{equation}
from the matter contribution to $H[N]$, pressure
\begin{equation}
 p_{\rm E}=-\frac{1}{N}\frac{\delta H_{\rm matter}[N]}{\delta
   \sqrt{\det h}}
\end{equation}
and stress
\begin{equation} \label{SE}
 S_{\rm E}^{ab}= -\frac{2}{N\sqrt{\det h}} 
\frac{\delta H_{\rm matter}[N]}{\delta h_{ab}}
\end{equation}
from derivatives of $H_{\rm matter}[N]$ by the metric, and energy and momentum
fluxes
\begin{equation} \label{JE}
 J^{\rm E}_a= \frac{1}{\sqrt{\det h}} \frac{\delta D_{\rm matter}[N^a]}{\delta
   N^a}
\end{equation}
in $D_{\rm matter}[N^a]$. (See again \cite{CUP}.) As indicated by the labels
``E'', these are stress-energy components as measured by Euclidean observers
whose worldlines are normal to equal-time hypersurfaces, or co-moving
observers in cosmological terminology. Deformed hypersurface-deformation
algebras therefore help in identifying physical deformations of fundamental
symmetries.

In the linear limit, (\ref{HHbeta}) gives rise to a deformed algebra in which
commutators between boosts and time translations, which both follow from
$H[N]$, are modified. Commutators in which spatial translations or rotations
appear, referring to $D[N^a]$, are undeformed. This behavior is in contrast to
the usual representation of the $\kappa$-Poincar\'e algebra
\cite{KappaPoincare,KappaPoincare2} in which only the commutator of boosts with
spatial translations is deformed,\footnote{Recall that $\{\cdot,\cdot\}$ is a
  Poisson bracket, not an anticommutator.}
\begin{equation}
 \{B_j,P_k\}= \delta_{jk} \left(\frac{1-\exp(-2\kappa P_0)}{2\kappa}+
   \frac{1}{2}\kappa \delta^{mn}P_mP_n\right)- \kappa P_jP_k\,.
\end{equation}
No such deformation can result from a gravitational theory which, like loop
quantum gravity, implements spatial diffeomorphisms unmodified. Moreover, the
$\kappa$-Poincar\'e algebra does not change the relations
\begin{equation}
 \{B_j,P_0\}=P_j \quad,\quad \{B_j,B_k\}= -\epsilon_{jkl}R_l
\end{equation}
which would be deformed in the linear limit of (\ref{HHbeta}); see
(\ref{BP}). The relations
\begin{equation}
\{P_{\mu},P_{\nu}\}=0 \quad,\quad \{R_j,P_0\}=0
\end{equation}
and
\begin{equation}
\{R_j,P_k\}=\epsilon_{jkl} P_l 
\quad,\quad \{R_j,R_k\}=\epsilon_{jkl} R_l
\quad,\quad \{R_j,B_k\}=\epsilon_{jkl} B_l
\end{equation} 
are undeformed in both cases.

The more-general parameterizations of generalized Poincar\'e algebras in
\cite{GeneralizedPoincare} allows a version in agreement with the deformation
found here. The ansatz made there leaves rotation generators undeformed, just
as we need it to make contact with our deformations. Boosts are deformed by a
parameterization of a boost operator deviating from
$\hat{x}_i\hat{p}_0-\hat{x}_0\hat{p}_i$. It follows from Eq.~(16) in
\cite{GeneralizedPoincare} that an undeformed commutator of a boost with a
spatial rotation, as implied by the unmodified (\ref{HD}), requires that only
the second part of the boost operator is modified, to
$\hat{x}_i\hat{p}_0-\beta\hat{x}_0\hat{p}_i$. No extra terms quadratic in
momenta and linear in position operators, as possible more generally, are
allowed. Qualitatively, the commutators (16) and (18) in
\cite{GeneralizedPoincare} then agree with the deformations obtained here in
the linear limit of (\ref{HHbeta}). However, while the deformed algebras of
observables agree, the required form of boost generators is not compatible
with the representation of $\kappa$-Minkowksi space discussed in
\cite{GeneralizedPoincare}.\footnote{We are grateful to Anna Pachol for
  pointing this out to us.} If a compatible representation can be found, it
would serve as a candidate for a quantum space-time model corresponding to a
deformed algebra (\ref{HHbeta}).

\section{Discussion}

Using the hypersurface-deformation algebra underlying canonical gravity, we
have extended deformed special relativity to deformed general relativity, with
several advantages:
\begin{itemize}
\item The deformation is derived from loop quantum gravity, which in recent
  years has produced several mutually consistent versions of deformed
  off-shell constraint algebras. The derivation clearly shows how quantum
  geometry gives rise to deformed fundamental symmetries. 
\item The generators of deformed hypersurface-deformation algebras have
  unambiguous physical meaning as boundary observables or stress-energy
  components. The deformation does not depend on one's choice of generators.
\item The deformation originally appears in position space, on which
  the classical structure functions of the algebra are defined. A
  corresponding momentum-space version exists only in certain regimes and
  depends more sensitively on which observer one refers to. Deformed special
  relativity, by contrast, is formulated in momentum space, and it is not only
  difficult to transform to position space but also problematic because of
  locality problems \cite{NonLocal1,NonLocal2,NonLocal3}.
\item Deformations of the hypersurface-deformation algebra in off-shell loop
  quantum gravity depend on the discreteness scale of quantum space or on the
  local extrinsic curvature, not on the mass of macroscopic bodies relative to
  the Planck mass. The ``soccer-ball problem'' does not occur, in a way that
  resembles proposed solutions in deformed special relativity
  \cite{DSRSoccer}.
\end{itemize}

Given these promising indications, the task of completing the understanding of
off-shell algebras in loop quantum gravity receives increased prominence. All
derivations so far have shown that the form (\ref{HHbeta}) of deformed
algebras and their deformation functions $\beta$ appears to be universal, but
they remain incomplete. Especially the inclusion of full holonomies with
integrations of the gravitational connection along curves remains a
challenge. The integrations involved may suggest non-local deformations of
the Poincar\'e algebra in the linear limit.

The relation between deformed special relativity and deformed general
relativity is not very direct. Nevertheless, the more-general viewpoint
developed here supports the basic ideas and motivations behind deformed
special relativity, even if there are differences in concrete
realizations. Some of the problems discussed extensively in deformed special
relativity do not appear in deformed general relativity or can easily be
solved, but a full analysis may still reveal new issues, including
observational ones. One may hope that the detailed methods developed and used
to scrutinize deformed special relativity can be applied to deformed general
relativity to put stringent tests on the underlying theory of loop quantum
gravity. Even though the derivation from loop quantum gravity makes use of
several assumptions and approximations, the broad consensus and universality
reached by all existing computations of off-shell constraint algebras, be it
by effective or operator methods, shows that loop quantum gravity can be ruled
out if its version of deformed general relativity is ruled out.

\section*{Acknowledgements}

We thank Anna Pachol and Lee Smolin for discussions.  This work was supported
in part by NSF grant PHY0748336.

%\bibliographystyle{../preprint}
%\bibliography{../Bib/QuantGra}

\begin{thebibliography}{10}

\bibitem{DSR1}
G.\ Amelino-Camelia,
\newblock Relativity: Special treatment,
\newblock {\em Nature} 418 (2002) 34--35

\bibitem{DSR2}
J.\ Magueijo and L.\ Smolin,
\newblock Lorentz Invariance with an Invariant Energy Scale,
\newblock {\em Phys.\ Rev.\ Lett.} 88 (2002) 190403

\bibitem{DSR}
J.\ Kowalski-Glikman,
\newblock Introduction to Doubly Special Relativity,
\newblock {\em Lect.\ Notes Phys.} 669 (2005) 131--159, [hep-th/0405273]

\bibitem{DSRSoccer}
F.\ Girelli and E.~R.\ Livine,
\newblock Physics of Deformed Special Relativity: Relativity Principle
  revisited, [arXiv:gr-qc/0412004]

\bibitem{DeformedQG}
Quantum symmetry, the cosmological constant and Planck scale phenomenology,
\newblock {\em Class.\ Quant.\ Grav.} 21 (2004) 3095--3110, [hep-th/0306134]

\bibitem{TQFTQG}
L.\ Smolin,
\newblock Linking Topological Quantum Field Theory and Nonperturbative Quantum
  Gravity,
\newblock {\em J.\ Math.\ Phys.} 36 (1995) 6417--6455, [gr-qc/9505028]

\bibitem{LorentzBianchi}
A.\ Dapor, J.\ Lewandowski, and Y.\ Tavakoli,
\newblock Lorentz Symmetry in QFT on Quantum Bianchi I Space-Time,
\newblock {\em Phys.\ Rev.\ D} 86 (2012) 064013, [arXiv:1207.0671]

\bibitem{DiracHamGR}
P.~A.~M.\ Dirac,
\newblock The theory of gravitation in Hamiltonian form,
\newblock {\em Proc.\ Roy.\ Soc.\ A} 246 (1958) 333--343

\bibitem{CUP}
M.\ Bojowald,
\newblock {\em Canonical Gravity and Applications: Cosmology, Black Holes, and
  Quantum Gravity},
\newblock Cambridge University Press, Cambridge, 2010

\bibitem{LapseGauge}
J.~M.\ Pons, D.~C.\ Salisbury, and L.~C.\ Shepley,
\newblock Gauge transformations in the Lagrangian and Hamiltonian formalisms of
  generally covariant theories,
\newblock {\em Phys.\ Rev.\ D} 55 (1997) 658--668, [gr-qc/9612037]

\bibitem{BrownYork}
J.~D.\ Brown and J.~W.\ York,
\newblock Quasilocal energy and conserved charges derived from the
  gravitational action,
\newblock {\em Phys.\ Rev.\ D} 47 (1993) 1407--1419

\bibitem{ReggeTeitelboim}
T.\ Regge and C.\ Teitelboim,
\newblock Role of surface integrals in the Hamiltonian formulation of general
  relativity,
\newblock {\em Ann.\ Phys.} 88 (1974) 286--318

\bibitem{ConsAlgebroid}
C.\ Blohmann, M.~C.\ Barbosa~Fernandes, and A.\ Weinstein,
\newblock Groupoid symmetry and constraints in general relativity. 1:
  Kinematics, [arXiv:1003.2857]

\bibitem{Regained}
S.~A.\ Hojman, K.\ Kucha\v{r}, and C.\ Teitelboim,
\newblock Geometrodynamics Regained,
\newblock {\em Ann.\ Phys.\ (New York)} 96 (1976) 88--135

\bibitem{LagrangianRegained}
K.~V.\ Kucha\v{r},
\newblock Geometrodynamics regained: A Lagrangian approach,
\newblock {\em J.\ Math.\ Phys.} 15 (1974) 708--715

\bibitem{Action}
M.\ Bojowald and G.~M.\ Paily,
\newblock Deformed General Relativity and Effective Actions from Loop Quantum
  Gravity,
\newblock {\em Phys.\ Rev.\ D} 86 (2012) 104018, [arXiv:1112.1899]

\bibitem{ConstraintAlgebra}
M.\ Bojowald, G.\ Hossain, M.\ Kagan, and S.\ Shankaranarayanan,
\newblock Anomaly freedom in perturbative loop quantum gravity,
\newblock {\em Phys.\ Rev.\ D} 78 (2008) 063547, [arXiv:0806.3929]

\bibitem{LTBII}
M.\ Bojowald, J.~D.\ Reyes, and R.\ Tibrewala,
\newblock Non-marginal LTB-like models with inverse triad corrections from loop
  quantum gravity,
\newblock {\em Phys.\ Rev.\ D} 80 (2009) 084002, [arXiv:0906.4767]

\bibitem{QSDV}
T.\ Thiemann,
\newblock {QSD V}: Quantum Gravity as the Natural Regulator of Matter Quantum
  Field Theories,
\newblock {\em Class.\ Quantum Grav.} 15 (1998) 1281--1314, [gr-qc/9705019]

\bibitem{InvScale}
M.\ Bojowald,
\newblock Inverse Scale Factor in Isotropic Quantum Geometry,
\newblock {\em Phys.\ Rev.\ D} 64 (2001) 084018, [gr-qc/0105067]

\bibitem{JR}
J.~D.\ Reyes,
\newblock {\em Spherically Symmetric Loop Quantum Gravity: Connections to
  2-Dimensional Models and Applications to Gravitational Collapse},
\newblock PhD thesis, The Pennsylvania State University, 2009

\bibitem{ScalarHol}
T.\ Cailleteau, J.\ Mielczarek, A.\ Barrau, and J.\ Grain,
\newblock Anomaly-free scalar perturbations with holonomy corrections in loop
  quantum cosmology,
\newblock {\em Class.\ Quant.\ Grav.} 29 (2012) 095010, [arXiv:1111.3535]

\bibitem{ThreeDeform}
A.\ Perez and D.\ Pranzetti,
\newblock On the regularization of the constraints algebra of Quantum Gravity
  in $2+1$ dimensions with non-vanishing cosmological constant,
\newblock {\em Class.\ Quantum Grav.} 27 (2010) 145009, [arXiv:1001.3292]

\bibitem{TwoPlusOneDef}
A.\ Henderson, A.\ Laddha, and C.\ Tomlin,
\newblock Constraint algebra in LQG reloaded : Toy model of a ${\rm U}(1)^{3}$
  Gauge Theory I, [arXiv:1204.0211]

\bibitem{TwoPlusOneDef2}
A.\ Henderson, A.\ Laddha, and C.\ Tomlin,
\newblock Constraint algebra in LQG reloaded : Toy model of an Abelian gauge
  theory - II Spatial Diffeomorphisms, [arXiv:1210.3960]

\bibitem{AnoFreeWeak}
C.\ Tomlin and M.\ Varadarajan,
\newblock Towards an Anomaly-Free Quantum Dynamics for a Weak Coupling Limit of
  Euclidean Gravity, [arXiv:1210.6869]

\bibitem{SigChange}
J.\ Mielczarek,
\newblock Signature change in loop quantum cosmology, [arXiv:1207.4657]

\bibitem{Rov}
C.\ Rovelli,
\newblock {\em Quantum Gravity},
\newblock Cambridge University Press, Cambridge, UK, 2004

\bibitem{ThomasRev}
T.\ Thiemann,
\newblock {\em Introduction to Modern Canonical Quantum General Relativity},
\newblock Cambridge University Press, Cambridge, UK, 2007, [gr-qc/0110034]

\bibitem{EffAc}
M.\ Bojowald and A.\ Skirzewski,
\newblock Effective Equations of Motion for Quantum Systems,
\newblock {\em Rev.\ Math.\ Phys.} 18 (2006) 713--745, [math-ph/0511043]

\bibitem{HigherTime}
M.\ Bojowald, S.\ Brahma, and E.\ Nelson,
\newblock Higher time derivatives in effective equations of canonical quantum
  systems,
\newblock {\em Phys.\ Rev.\ D} 86 (2012) 105004, [arXiv:1208.1242]

\bibitem{HigherCurvHam}
N.\ Deruelle, M.\ Sasaki, Y.\ Sendouda, and D.\ Yamauchi,
\newblock Hamiltonian formulation of $f({\rm Riemann})$ theories of gravity,
\newblock {\em Prog.\ Theor.\ Phys.} 123 (2009) 169--185, [arXiv:0908.0679]

\bibitem{SphSymm}
M.\ Bojowald,
\newblock Spherically Symmetric Quantum Geometry: States and Basic Operators,
\newblock {\em Class.\ Quantum Grav.} 21 (2004) 3733--3753, [gr-qc/0407017]

\bibitem{SphSymmHam}
M.\ Bojowald and R.\ Swiderski,
\newblock Spherically Symmetric Quantum Geometry: Hamiltonian Constraint,
\newblock {\em Class.\ Quantum Grav.} 23 (2006) 2129--2154, [gr-qc/0511108]

\bibitem{InhomLattice}
M.\ Bojowald,
\newblock Loop quantum cosmology and inhomogeneities,
\newblock {\em Gen.\ Rel.\ Grav.} 38 (2006) 1771--1795, [gr-qc/0609034]

\bibitem{CosConst}
M.\ Bojowald,
\newblock The dark side of a patchwork universe,
\newblock {\em Gen.\ Rel.\ Grav.} 40 (2008) 639--660, [arXiv:0705.4398]

\bibitem{KappaPoincare}
J.\ Lukierski, H.\ Ruegg, A.\ Nowicki, and V.~N.\ Tolstoi,
\newblock Q deformation of Poincare algebra,
\newblock {\em Phys.\ Lett.\ B} 264 (1991) 331--338

\bibitem{KappaPoincare2}
J.\ Lukierski and H.\ Ruegg,
\newblock Quantum kappa Poincare in any dimension,
\newblock {\em Phys.\ Lett.\ B} 329 1994, [hep-th/9310117]

\bibitem{GeneralizedPoincare}
D.\ Kovacevic, S.\ Meljanac, A.\ Pachol, and R.\ Strajn,
\newblock Generalized Poincare algebras, Hopf algebras and kappa-Minkowski
  spacetime,
\newblock {\em Phys.\ Lett.\ B} 711 (2012) 122--127, [arXiv:1202.3305]

\bibitem{NonLocal1}
S.\ Hossenfelder,
\newblock Bounds on an energy-dependent and observer-independent speed of light
  from violations of locality,
\newblock {\em Phys.\ Rev.\ Lett.} 104 (2010) 140402, [arXiv:1004.0418]

\bibitem{NonLocal2}
G.\ Amelino-Camelia, M.\ Matassa, F.\ Mercati, and G.\ Rosati,
\newblock Taming nonlocality in theories with Planck-scale-deformed Lorentz
  symmetry,
\newblock {\em Phys.\ Rev.\ Lett.} 106 (2011) 071301, [arXiv:1006.2126]

\bibitem{NonLocal3}
S.\ Hossenfelder,
\newblock Reply to arXiv:1006.2126 by Giovanni Amelino-Camelia et al,
  [arXiv:1006.4587]

\end{thebibliography}

\end{document}